\documentclass[conference]{IEEEtran}

\usepackage[hyphens]{url}
\usepackage[hidelinks]{hyperref}
\usepackage{graphicx}

\ifCLASSINFOpdf
\else
  \usepackage{graphicx}
\fi
\usepackage{color, soul}
\usepackage{graphicx}
\usepackage{paralist}
\usepackage[tight,footnotesize]{subfigure}
\usepackage{caption}
\usepackage{xcolor}
\let\labelindent\relax
\usepackage[inline]{enumitem}
\usepackage{amsmath} 
\usepackage[nolist,nohyperlinks]{acronym}
\usepackage{comment}

\begin{acronym}
    \acro{TLS}{Transfer Layer Security}
    \acro{DAPS}{Dynamic Attribute Provisioning Service}
    \acro{DAT}{Dynamic Attribute Token}
    \acro{IDS}{International Data Spaces}
    \acro{IDSA}{International Data Spaces Association}
    \acro{MEC}{Multi-Access Edge Computing}
    \acro{IoT}{Internet of Things}
   
\end{acronym}

\usepackage{pgfplots}
\usepackage{pgfplotstable}
\usetikzlibrary{patterns}
\begin{document}
\title{EdgeDS: Data Spaces enabled \\Multi-access Edge Computing}

\author{\IEEEauthorblockN{Ioannis Kalogeropoulos, Maria Eleftheria Vlontzou,\\Nikos Psaromanolakis, Eleni Zarogianni, Vasileios Theodorou}
\IEEEauthorblockA{Intracom Telecom, Greece\\
Email: \{ikalogerop, mevlon, nikpsarom, ezarog, theovas\}@intracom-telecom.com}}


%Telephone: (800) 555--1212, Fax: (888) 555--1212}


\maketitle

\begin{abstract}
%\boldmath
The potential of Edge Computing technologies is yet to be exploited for multi-domain, multi-party data-driven systems. One aspect that needs to be tackled for the realization of envisioned open edge Ecosystems, is the secure and trusted exchange of data services among diverse stakeholders. In this work, we present a novel approach for integrating mechanisms for trustworthy and sovereign data exchange, into Multi-access Edge Computing (MEC) environments. To this end, we introduce an architecture that extends the ETSI MEC Architectural Framework with artifacts from the International Data Spaces Reference Architecture Model, accompanied by processes that automatically enrich Edge Computing applications with data space capabilities in an as-a-service paradigm. To validate our approach, we implement an open-source prototype solution and we conduct experiments that showcase its functionality and scalability. To our knowledge, this is one of the first concrete architectural specifications for enabling data space features in MEC systems.

\end{abstract}


% \fi
\IEEEpeerreviewmaketitle

\section{Introduction}

\begin{figure*}[htbp!]
\begin{centering}
%\resizebox{.87\textwidth}
  \includegraphics[width=0.7\textwidth]{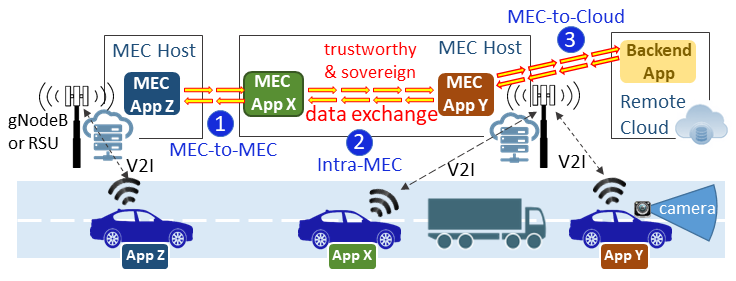}
\caption{Motivating Use Case from the Autonomous Driving domain}
\label{fig-motivating-UC}
\end{centering}
\end{figure*}

Edge Computing, i.e., the Cloud Computing paradigm that brings data processing and data storage in close proximity to---or directly on---the ``edge’’ network nodes of data providers and end users, is increasingly gaining traction. The promise of support for next-generation decentralized applications and services through unprecedented optimizations in delays and bandwidth usage, has already established Edge Computing as a fundamental pillar of the 5G ecosystem, and nowadays is being recognized, accompanied with Artificial Intelligence, as a key enabler for future 6G networks.

At the same time, applications are becoming more distributed in nature and the value chain of data services is crossing multiple administrative and business domains to offer competitive advantages to data-driven ecosystem stakeholders. The communications and infrastructures domains appear to follow this trend, by opting for more open and interoperable distributed architectures that can minimize CapEx and OpEx costs and catalyze the rollout and evolution of advanced modern services. Open Radio Access Network (Open RAN) and Service Based Architecture (SBA) of the mobile core are characteristic examples of this new model in the telecommunications domain, indicating that the management, automation and optimization of network and application services will shortly not be performed within isolated administrative environments, but rather in a synergistic operation of collaborating systems.

To showcase the benefits of data-driven collaboration across systems and actors, we employ a motivating use-case example from the Autonomous Driving domain, as illustrated in Fig.~\ref{fig-motivating-UC}. We envision different scenarios for autonomous, connected vehicles on a highway, which are powered with advanced safety, traffic-routing and other features delivered by the enactment of data service ecosystems. For instance, we consider the `see-through' case of a vehicle with an intention to overtake a track and is being temporarily provided access to a video stream of a car, equipped with a camera, preceding that track. Similar scenarios are applicable, such as the `platooning' of vehicles that can drive in a group-like coordinated manner, sharing sensing information, and the exchange of safety or traffic alerts among vehicles on the highway. A particularly interesting aspect among those scenarios is the heterogeneity of provided manufacturer technologies, the connectivity modes (e.g., cellular vehicle-to-infrastructure (V2I) connectivity via gNodeBs or access through road-side units (RSU) by the highway operator), the data application providers and operators, the various tiers of distributed applications (e.g., client tier -- MEC tier -- backend tier) as well as the co-existence of legacy, non-connected vehicles to the highway.

In an attempt to encompass such use cases of interaction at the Network Edge, the ETSI Multi-access Edge Computing (MEC) ISG has introduced a comprehensive Architectural Framework\cite{mec} that facilitates not only the life-cycle management of applications (i.e., MEC Applications) that are running on virtualization Edge infrastructures, but also the creation of an ecosystem for the exchange of Edge Services among MEC Applications. Despite a clear definition of functional blocks and workflows for the establishment of service exchange mechanisms, this framework is by-design application-agnostic and does not specify in detail the security and authentication primitives that would allow cross-domain, multi-party collaboration between edge actors. Nevertheless, we argue that the establishment of trustworthy, secure and data-resource-aware data exchange between MEC Applications is pivotal to the opening of fragmented edge computing environments and to unleashing the potential of open, decentralized architectures.

Recently, several ``Data Space'' initiatives have emerged, specifically focused on the definition standards and procedures to ensure the reliable, trustworthy and sovereign exchange of data services, across organizational boundaries, using well-defined, open interfaces. Although the added value of enhanced data security in Edge Computing  Industry 4.0 applications is well identified \cite{edge-data}, straightforward and practical instructions of how such merging of paradigms should take place are still missing. 

In this work, we take an important step of incorporating Data Spaces features as native services in MEC systems. To this end, we propose an extension of the ETSI MEC architecture, by introducing an ``IDS-Connector-as-a-service'' approach, instilled directly into the MEC mechanisms. We present in detail the architecture to support such features, as well as the workflow steps for the data spaces-enabled interaction of MEC Applications. To our knowledge, our work is the first to offer concrete directives towards the realization of data spaces-enabled Edge Environments and in this direction, we also provide a prototype implementation of our approach as open source code, with the intention to foster more research by the community in this area.   

The rest of the paper is organized as follows. Section~\ref{sec: background} provides background information on data spaces, MEC and early approaches on the intersection of Edge Computing and Data Spaces. Subsequently, Section~\ref{edgeds} presents the architectural view of our approach detailing its MEC Plaform automation mechanisms. Section~\ref{sec:evaluation} describes our experiments from the application of our implemented prototype on different scenarios and finally, the paper ends with a discussion of our findings and concluding remarks in Section~\ref{sec:discussion} and Section~\ref{sec:conclusion} respectively.

\section{Background}
\label{sec: background}
% In which context have the data spaces been used
\subsection{International Data Spaces}
The notion of a \emph{data space} is not new. Moving from traditional Database management systems, data spaces offer enhanced capabilities for browsing through catalogs, local storage and index, and advanced search and query mechanisms \cite{DS}. Expanding on this notion lay data spaces initiatives that have recently spawned, such as the \ac{IDSA}\footnote{\url{https://internationaldataspaces.org/}}. The main objective of data spaces involve the secure and trusted data exchange among stakeholders, whilst ensuring data sovereignty and monitoring capabilities for the entire data workflow.

\ac{IDS} \cite{IDSA} represent a decentralized data sharing architectural concept, in which data physically remain at their source and only transferred to another interested part once data exchange requests are instantiated. Data sovereignty and trust are established, since each participant is able to attach usage restrictions to their data and monitor data transactions through continuous monitoring and logging. Additionally, security is ensured, through the identity evaluation of each participant by \ac{IDS} certified bodies. Furthermore, \ac{IDS} offers data processing capabilities through certified  services, metadata storing, as well as metadata-query functionalities that enable participants to search for the appropriate data sources and request access to the respective data.

Among core participants in the IDS ecosystem are the \emph{data provider} and the \emph{data consumer}. The data provider is the entity that provides access to a data source, and attach respective usage restrictions, while the data consumer is the participant, who can search for appropriate data sources and after accepting usage policies set by the provider, can obtain access to the data.

A central component of the \ac{IDS} is the \ac{IDS} Connector,  which enables the data exchange between data providers and consumers. Each of them is represented by a connector, which allows the registration of offered data resources, along with the metadata that describe them. Additionally, the connector facilitates the attachment of usage rules to the data on the provider's side and the usage contract negotiation on the consumer's part, which ultimately leads to the bilateral agreement between involved parties, and ensures the enforcement of data access policies. 

Security, being a strategic requirement of the IDS \cite{IDSA}, is based on the certification and dynamic monitoring of all participants and technical components (e.g. connectors), as well as the \ac{TLS} protected communication between connectors. Providers and consumers should be successfully certified to participate in the IDS along with their certified IDS connectors. If these conditions are met, a unique \ac{IDS}-ID is generated and a digital certificate (X.509) is issued for the participant-connector combination, thus enabling the identification, authentication and point-to-point encryption for the communication between connectors. The connector can then be registered at the \ac{DAPS} component and request a \ac{DAT} through which the validity of the connector’s self-description is certified. The \ac{DAT} is included in every outgoing communication message of the connectors, thus ensuring the trustfulness of communication partners at any time.

The \ac{IDS} framework has recently gained traction and has been identified as a data ecosystem enabler across many industry domains, including the Energy, Manufacturing and Health Care data sectors \cite{digital-twins}, providing a secure, sovereign and trustworthy framework for data sharing. Noteworthy initiatives, such as the Catena-X project \cite{Catena} and the Mobility Data Space \cite{Mobility}, are exploiting IDS components for the development of their data space ecosystems; however these projects are work in progress and not mature enough to be considered for production. 

%However, ...

\subsection{MEC Architectural Framework}
\ac{MEC}, as proposed by ETSI \cite{mec}, is a highly promising framework that paves the way towards satisfying ultra low-latency requirements, as well as, providing rich computing environment for value-added services closer to end-users. Specifically, \ac{MEC} enables the implementation of \ac{MEC} applications as software-only entities, existing on top of a Virtualisation infrastructure, which is located on or close to the network edge. This framework defines a reference architecture comprised of various entities, acting at a system-, host- and network-level, remaining however generic enough to allow for the development of extensions. 

The core functionalities of the \ac{MEC} architecture are realized at the \ac{MEC}-host level, which contains the \ac{MEC} platform and a Virtualisation infrastructure, which provides computing, storage and network resources for the \ac{MEC} applications. The \ac{MEC} Platform provides an environment, where \ac{MEC} applications can discover, advertise, consume and offer \ac{MEC} services, while maintaining responsibility for receiving various traffic rules and DNS. Furthermore, the \ac{MEC} Platform is responsible for offering its own \ac{MEC} services, regarding the management of \ac{MEC} Applications services and location information about the registered Applications, Zones and Access Points, etc. Moreover, \ac{MEC} applications exist as Virtualised applications on top of the Virtualisation infrastructure provided by the \ac{MEC} host and are able to communicate with other applications towards consuming or providing services.

The process of discovering and utilizing a \ac{MEC} service by a \ac{MEC} App is well defined. Specifically, \ac{MEC} Applications perform \textit{availability queries} to the \ac{MEC} Platform and receive a list of all the available \ac{MEC} services, along with the necessary information required for their consumption. Subsequently, each \ac{MEC} App possesses the exposed APIs for the desired MEC service and can access the data provided by that service. 

Several works \cite{mec-extensions} have proposed extensions of the MEC architecture with complementary technologies and architectures (by modifying or adding new entities or reference points), such as Network Functions Virtualization (NFV)\cite{mano}, Software Defined Networking (SDN), and Cloud-Radio Access Network (C-RAN)\cite{mec-extensions2}, thus expanding MEC capabilities for improved traffic management, task offloading and resource orchestration and virtualization across edge nodes, among others.

% Data spaces on (or relevant) edge platforms
\subsection{Data Spaces ``on the Edge''}
Several studies have explored the idea of incorporating data spaces to edge computing architectures, however, to the best of our knowledge, none has gone thus far as to implement them in practice and even utilize a specific reference architecture, such as IDS \cite{IDSA}.Trakadas et al. \cite{hybrid-clouds} proposed a decentralized hybrid cloud \ac{MEC} architecture  and highlighted key challenges that emerge in hybrid clouds for data-intensive \ac{IoT} applications, such as issues in privacy and security, which on a conceptualization level proposed to be tackled by utilizing data spaces, to ensure secure and trusted data exchange and provide distributed identity management. 
Zeiner et al. \cite{time-aware} highlighted the need for data sharing mechanisms between neighboring edge servers and proposed the concept of time-aware data spaces, as a computing unit for collecting and analyzing data, while also ensuring the validity of the data. Sun et al. \cite{IoT-privacy} designed an \ac{IoT} data sharing privacy-preserving model that is based on the edge computing service, and establishes the virtual data management service, by using a data space layer for the acquisition, query and analysis of data, which are physically distributed in multiple systems.

Although the aforementioned approaches highlight the advantages of incorporating Data Space functionalities on edge platforms, there is no other specific architectural framework or detailed workflows proposed to indicate how this could be realized. Furthermore, none of these approaches propose the integration of specific Data Space frameworks, such as the \ac{IDS} or reference Edge architectures, such as ETSI \ac{MEC}. The key contribution of this work is the integration of \ac{IDS} into the \ac{MEC} Architecture and the use of the \ac{IDS} Connector component for the communication and data exchange among \ac{MEC} Applications or the \ac{MEC} Platform by proposing the concept of the \ac{IDS} Connector-as-a-Service. With this new functionality, \ac{MEC} hosts can provide a trusted and secure data sharing environment among the \ac{MEC} Applications or \ac{MEC} Platform, where data sovereignty is preserved, by enabling the attachment of strictly defined, uniform access restrictions to the data.

\section{The EdgeDS Approach}\label{edgeds}

\subsection{Architecure}
\begin{figure}[htbp!]
\begin{centering}
\includegraphics[width=0.49\textwidth]{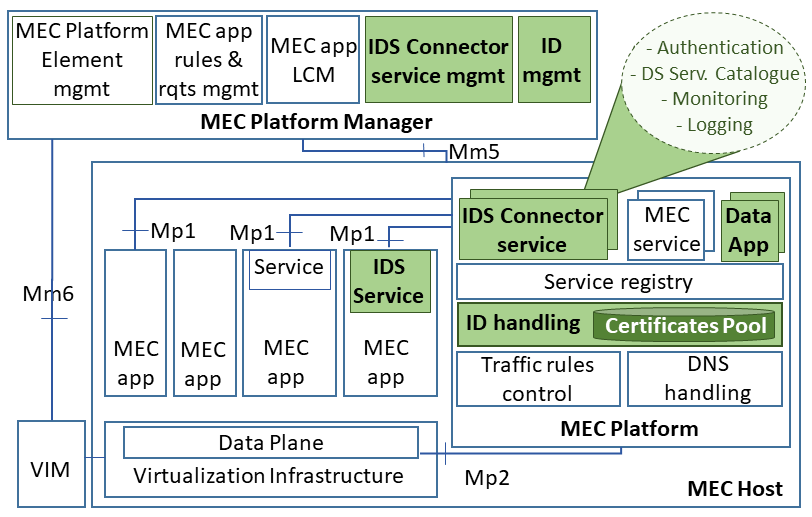}
\caption{ Extension of ETSI MEC Architectural Framework with Data Spaces Enablement.}
\label{fig:proposed_architecture}
\end{centering}
\end{figure}

In this work, we suggest the integration of Data Spaces concepts with \ac{MEC} to exploit on the secure and trustworthy data sharing mechanisms. Our introduced architecture is depicted in Fig.~\ref{fig:proposed_architecture}, where extensions to the original ETSI MEC Architecture are denoted as additional components with highlighted (green) background. Specifically, as shown in Fig. \ref{fig:proposed_architecture}, each \ac{MEC} Application (MEC App) can act as a data provider/consumer, in accordance with \ac{IDS} roles distinction. By introducing the concept of \emph{IDS Connector-as-a-Service} among the functionalities provided by the \ac{MEC} Platform, each application is capable of obtaining its own connector---which exists within the \ac{MEC} Platform as a MEC Platform-offered service instance---and exchange data with any application that has a registered connector inside or outside the \ac{MEC} host.

The desired architecture is able to guarantee the secure transfer of data between two applications, respecting IDS-defined constraints, regardless of the type of devices or the network these devices are connected to. Hence, six distinct use cases exist that can be defined, depending on the type of entities that participate in the data exchange scheme, as well as their location, relative to a MEC Host. In particular, the proposed architecture supports the consumption of data services between the \ac{MEC} Platform and a \ac{MEC} Application or between two \ac{MEC} Applications, while any of those exchanging parties can belong in the same or different \ac{MEC} Host(s). Moreover, data exchange between an External App and a \ac{MEC} Platform or a \ac{MEC} Application is also supported.

% Explain the MEC services/ catalogs
The integration of IDS policies inside the MEC architecture paves the ground for a straightfoward exchange of data between any of the two aforementioned entities. As proposed both in \ac{MEC} and \ac{IDS} architectures, a list (catalog) of the registered data services is made available. Specifically, the former defines a list of the available services provided by each \ac{MEC} Application, while the latter proposes one or more catalogues of the available data resources within each connector. Keeping in line with both approaches, a list of all the available \ac{MEC} Services that are not \ac{IDS} enabled is reserved, along with all the connectors (i.e., connector service instances) present within the MEC Host. This approach supports the concept of Connector-as-a-Service, providing to \ac{MEC} Applications an interface of the connectors similar with the one of the regular \ac{MEC} services. Additionally, the information relevant to the data provided through the connectors is accessible within each connector's catalogue(s).

Furthermore, the proposed integration of IDS into the MEC architecture enables the incorporation of specific IDS-certified \emph{Data Apps} to the MEC Platform. These applications can be regarded as data processing services (Extract-Transform-Load (ETL), Analytics, ML models, etc), which can be used in the data exchange workflow and are either offered by the Provider’s, the Consumer’s or third-party IDS Connectors. Data Apps are made available to the MEC Platform, by being registered at a specific IDS component (App Store) within the MEC Platform Service Registry. This allows for data interactions between MEC Applications, focusing on the exchange of data processing results. For instance, with respect to the example from the Autonomus Driving domain described in Section~1, an Object Detection IDS-certified Data App could be offered by a third-party in the MEC Platform, enabling a MEC Application to send certain image frames and another MEC Application to retrieve the detected objects from those images, rather than the images themselves. Apart from offering rich data processing utilities at the Edge, data apps provide also an additional means of privacy-preservation, since they enable the exchange of data processing output and aggregated data, instead of exchanging raw datasets among MEC Apps. 

% Analyze MEC1App2 - MEC2App1
In the scenario of two \ac{MEC} Applications exchanging data, we denote as \ac{MEC} App2 the application that attempts to receive data and as \ac{MEC} App1 the one that provides the desired data. After the completion of instantiation for both applications and the assignment of a certified IDS connector to each of them (more details in Subsection \ref{automation} below), the connectors' information are made available though the \ac{MEC} Platform's relevant API for service discovery and availability. While data services that are not \ac{IDS} enabled are registered directly on the \ac{MEC} Platform, \ac{MEC} App1 registers the data resources it provides on its own connector. In order to consume the desired data services, \ac{MEC} App2 receives the information of \ac{MEC} App1's connector. Subsequently, MEC App2 performs a request to its own connector, encapsulating the information of the other connector, along with the identifiers of the targeted resources. Finally, if the rule policies attached to the contract of the requested resource clear the request, then data transfer from the provider to the consumer is followed as defined by \ac{IDS}.

Regarding the other use cases, in the scenario of a \ac{MEC} Application consuming data services from the \ac{MEC} Platform of its own host, the process is the same as above, since the application is able to discover Platform's services through the same APIs, while the relevant connectors will be located within the same host. The process to be followed, however, when the two parties of the transfer belong to a different \ac{MEC} Host, needs to be extended so that both can access the information regarding the relevant connectors. In particular, as proposed by ETSI \cite{mec} they should be able to communicate through the Reference Point Mp3, which is reserved for accessing other \ac{MEC} Hosts.

% why we decided to put the connectors inside MEC Platform.
Towards incorporating the concept of Dataspaces and taking advantage of their features, we adopt the current version of IDS reference architecture as proposed by the IDSA, because it is generic enough to encompass diverse scenarios. Moreover, the placement of each connector within the \ac{MEC} Platform is motivated by the need of restricting them in a controlled environment of well-defined security and available resources. By offering the Connector-as-a-Service feature, we automatically augment  \ac{MEC} Applications with Dataspace capabilities, refraining from imposing on them any data-spaces-related extension requirements.

\subsection{MEC Platform Automation Mechanisms} \label{automation}
% workflow figure - automation
While the life-cycle management (LCM) of data spaces-enabling services, such as IDS connectors, into the Edge Cloud continuum, is quite challenging and toilsome, and also requires continuous interaction with different orchestration and management entities, this work offers automation into the context of composition, deployment, activation, authentication and authorization, as well as LCM of those services. Moreover, it also ensures scalability, high availability, and security of the system, through the usage of a MEC Platform Manager, which encapsulates relevant automation features.

\begin{figure}[htbp!]
\begin{centering}
\includegraphics[width=0.49\textwidth]{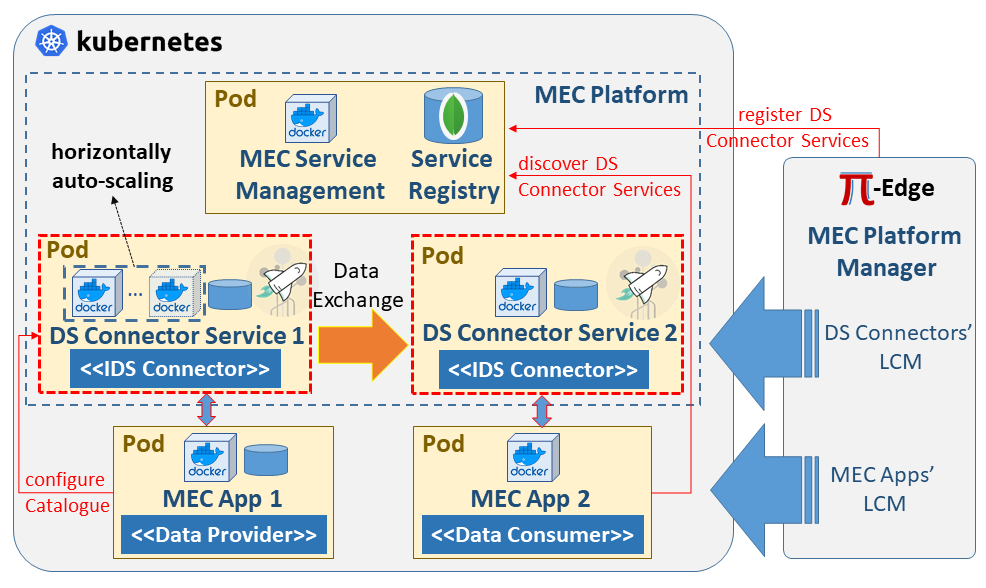}
\caption{ Experimental Setup.}
\label{fig:exp-setup}
\end{centering}
\end{figure}

To this end, as shown in Fig. \ref{fig:exp-setup}, through the adoption of the \emph{IDS Connector-as-a-Service} concept, the MEC Platform Manager is capable of composing, deploying, instantiating and managing the MEC Applications and IDS connectors that are running at the Edge, via its \emph{IDS Connector service management component}. To this end, we employ a simple high-level extension to the MEC Information Model (IM), by simply denoting through an optional boolean attribute whether a MEC app shall be data-spaces-enabled, and the rest is handled by our system automatically. In detail, the MEC Platform Manager undertakes a) the optimal deployment of MEC Applications and IDS connectors at the appropriate edge nodes, based on their available resources, b) the registration of IDS connectors as a service into the MEC Platform Service Registry, c) the network configuration and composition, in that it connects an IDS connector with a MEC App if the latter is denoted as data-spaces-enabled, and d) the assignment of a certificate to each IDS connector, during its instantiation, derived from a pool of available certificates that exist within the MEC Platform, as determined in IDS security strategic requirements analyzed in section \ref{sec: background}. For the latter functionality, we assume that each MEC Host is equipped with a set of pre-accredited certificates, as an output of relevant actions performed by the \emph{ID management} component of the \emph{MEC Platform Manager}, upon its interactions with centralized or distributed authorization entities. Apart from the LCM automation of those services and Applications, the MEC Platform is responsible for handling all necessary actions for the successful and efficient termination of a MEC App and its associated IDS connector. Hence, the termination step also includes the de-registration of the IDS connector service from the MEC Platform Service Registry, and the release of the associated certificate back to the certification pool.

% describe authentication

\section{Evaluation} \label{sec:evaluation}
% Describe experiment and setup 1
% testbed, pi-edge, mec platform, github (mec code, pi-edge container, documentation), tech stack (IDSA, databse, k8s, docker)
Our prototype implementation of the proposed architecture is depicted in Fig.~3 and it included the integration of \ac{IDS} architecture, implemented by Fraunhofer \footnote{\url{https://github.com/International-Data-Spaces-Association/DataspaceConnector}}, together with a \ac{MEC} Platform realisation and $\pi$-Edge, that was first introduced in \cite{piedge}, both implemented from scratch. Specifically, the latter is an orchestration and management platform for automating the LCM of Platform as a Service (PaaS) functions at the edge. Moreover, its functionalities were used to instantiate and register the \ac{MEC} Applications and the connectors. Besides the above technologies, Docker was used to containerize each module, and Kubernetes for their deployment. Moreover, $\pi$-Edge encompasses an internal MongoDB Database, while each IDS connector has its own PostgreSQL. For our experiments, the Edge infrastructure is comprised of an Intel(R) Xeon(R) Silver 4214 CPU @ 2.20GHz processor and 32 GB RAM. The code used for the implementation is made open and can be found at \href{https://github.com/jkalogero/EdgeDS}{https://github.com/jkalogero/EdgeDS}.

% We implement the architecture described in section \ref{edgeds} and evaluate its performance for transferring resources of different sizes. As a baseline scenario for comparison, we also conduct the data transfer via direct \ac{MEC} services.
% ... We conduct the same experiments between \ac{MEC} Applications but not under \ac{IDS} concept and compare our findings.

For evaluating the proposed architecture, experiments were conducted on transferring data of different sizes, ranging from 1 MB to 150 MB and were inspired from use cases similar to the one depicted in Figure \ref{fig-motivating-UC}. For both experiments, we recorded the time needed for the end-to-end process of instantiating and registering the connectors, creating the \ac{IDS} resources along with the catalogs, rule policies, contracts and all the necessary \ac{IDS} entities, fetching the \ac{MEC} App1' (provider) information, requesting the desired resources, and finally downloading the locally within MEC App2. As a baseline scenario for comparison, we have also conducted the data transfer via direct \ac{MEC} services.
% Specifically, as depicted in Figure \ref{fig:experiment}, we measure the time needed for 

\begin{figure}[htbp!]
\begin{tikzpicture}[
/pgfplots/every axis/.style={ % <- added /pgfplots/ 
    ybar stacked,
    title={Data Exchange Total Time},
    ymax=40,
    ymin=0, %<- removed ymin
    symbolic x coords={
      5, 10, 25, 50, 75, 100, 125, 150},
    bar width=3.5pt,
      legend style={font=\small, at={(0,1)},anchor=north west, draw=none},
    ylabel={Time(s)},
    xlabel={Data Size (MB)},    
    xtick=data,
    x tick label style={font=\small, 
    },
    point meta=rawy,
  },
]

\begin{axis}[bar shift=-6pt, bar width=10pt][legend style={nodes={scale=0.5, transform shape}}, ]
\addplot coordinates {(5,9.7) (10,9.7) (25,9.7) (50,9.7) (75,9.7) (100, 9.7) (125, 9.7) (150, 9.7)};\label{p1}
\addplot coordinates {(5,1.37) (10,1.37) (25,1.37) (50,1.37) (75,1.37) (100, 1.37) (125, 1.37) (150, 1.37)}; \label{p2}
\addplot coordinates {(5,1.788) (10,2.478) (25,5.492) (50,9.131) (75,11.7017) (100, 15.603) (125, 19.446) (150, 22.430)};\label{p3} 
\end{axis}
\begin{axis}[bar shift=6pt,hide axis, bar width=10pt][legend image post style={mark=*}] %<- increased
\addplot+[pattern=north east lines] coordinates {(5,3) (10,3) (25,3) (50,3) (75, 3) (100, 3) (125, 3) (150, 3)};\label{p4}
\addplot+[pattern=north east lines] coordinates {(5, 0.038) (10, 0.038) (25, 0.038) (50, 0.038) (75, 0.038) (100, 0.038) (125, 0.038) (150, 0.038)};\label{p5}
\addplot+[pattern=north east lines] coordinates {(5,0.088) (10,0.098) (25,0.192) (50,0.37) (75,0.4445) (100, 0.692) (125, 0.6997) (150,0.8255)};\label{p6}
\end{axis}
\node [draw,fill=white] at (rel axis cs: 0.13,-0.38) {\shortstack[l]{
\ref{p1} prepare IDS services \\
\ref{p2} configure IDS catalogue \\
\ref{p3} data exchange over\\ \hspace{6pt} IDS connectors }};
% Second "Legend" node
\node [draw,fill=white] at (rel axis cs: 0.79,-0.38) {\shortstack[l]{
\ref{p4} instantiate MEC Applications \\
\ref{p5} register MEC Service \\
\ref{p6} direct data exchange \\ \hspace{6pt} between MEC Applications }};
\end{tikzpicture}
\caption{ Data exchange for various data sizes.}
\label{fig:experiment}
\end{figure}
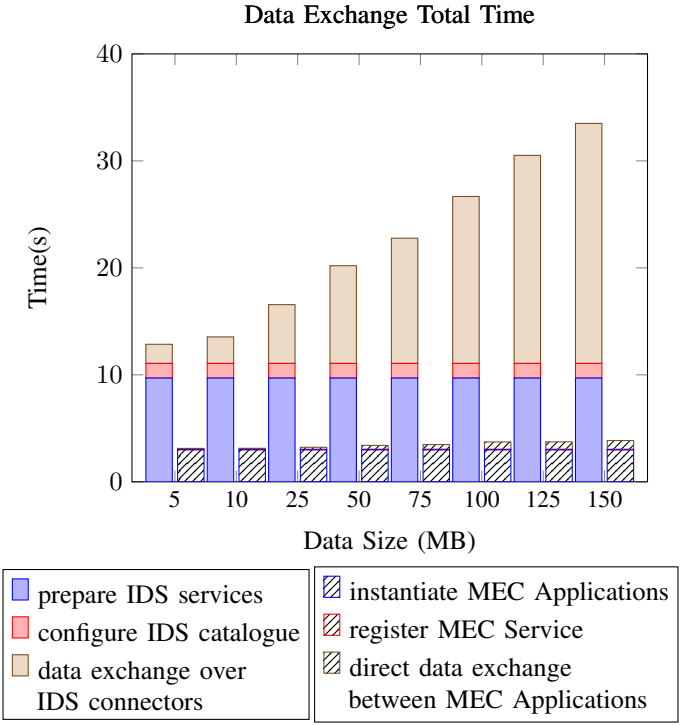

% Results

% Discuss results
As seen in Fig. \ref{fig:experiment},  the total time for a complete data transaction in the case of \ac{IDS}-enabled \ac{MEC} applications was greater compared to the \ac{MEC} applications direct communication scenario. Specifically, the time required for the \ac{IDS} services preparation step was three times greater than the instantiation time of \ac{MEC} Applications in the direct \ac{MEC} App communication scenario.
 
With regards to the IDS catalogue configuration step, the required time for the \ac{IDS} enabled scenario was stable independently of the data volume, and equal to 1.37s, whereas in the case of direct \ac{MEC} App communication, the required time to register the respective \ac{MEC} service was two orders of magnitude lower, and equal to 0.038s. 

Lastly, the data exchange time increased,  as a result of the volume of the exchanged data resource, and was also two orders of magnitude greater in the case of \ac{IDS} enabled \ac{MEC} Applications, compared to the direct \ac{MEC} Application data transfer scenario.  

% Describe experiment and setup 2
In order to conduct a comprehensive analysis of the scalability potential of the proposed architecture, we carried out a series of experiments to evaluate its capacity to manage a substantial volume of traffic, by subjecting the provider to numerous requests for data exchange. By concurrently generating multiple requests for data services from a single data provider, we were able to observe a significant increase in the pod's CPU utilization, as well as in the time required to complete the data exchange.

In order to tackle the aforementioned issue, we utilized the autoscaling feature, supported by the Kubernetes system. Upon conducting the same experiments, we observed the gradual provisioning of multiple additional connector pods to effectively handle the significant workload. Experiments with various intensive workloads, resulted to successful data transfers with a limited average CPU utilization, while significantly reducing the time required to complete the transfers. 
% Maybe add a final sentence saying the multiple mec app can concurrently consume and take advantage of these services

% * 
\section{Discussion} \label{sec:discussion}
During our evaluation, we have successfully deployed the integration of our introduced \ac{IDS} Connector-as-a-Service approach within the \ac{MEC} Architecture. Completion time of a data exchange process was used as a key performance indicator in our experiments, measuring the management overhead of introducing data space features to the MEC system.
Specifically, instantiation time in the \ac{IDS} Connector-as-a-Service scenario was found to be greater, as it entails the preparation of the two IDS connectors, and their registration to the \ac{MEC} Platform service registry, on top of the \ac{MEC} applications deployment. However, this step only takes place once and therefore does not impose any further delay in the data exchange process, in case of existing \ac{IDS} enabled \ac{MEC} applications. 

 In the direct \ac{MEC} App communication, registration of the \ac{MEC} service corresponds to the time needed for \ac{MEC} App1 to register the respective data service to the \ac{MEC} Platform, whereas for IDS-enabled \ac{MEC} applications, the configuration time of the \ac{IDS} catalogue involves the time \ac{MEC} App1 needs to obtain the service information of Connector 2 from the \ac{MEC} Platform and use the connector to register the offered data resource’s metadata and usage restrictions, which is realized through several API calls. Additional delay could also result from the continuous logging the Connector provides as a live monitoring feature.

In the direct \ac{MEC} App communication case, data exchange only depends on the time \ac{MEC} App2 requires to obtain the service information of \ac{MEC} App1 from the \ac{MEC} Platform, as well as to request and receive the data directly from \ac{MEC} App1. However, for the IDS-enabled scenario, this time corresponds to the retrieval of the two Connectors’ service information from the \ac{MEC} Platform by \ac{MEC} App2 and the time needed to request Connector 2 available data sources, negotiate the contract agreement based on the usage rules attached to the requested data resource and eventually access the data. Further delays could be caused by the fact that downloaded data are locally stored in the Connector's database as a bytestream and are automatically decoded on the API call, as well as by the continuous logging feature of the Connectors, which as mentioned above, enhances security and data sovereignty according to the \ac{IDS} protocol.

%A significant limitation of the IDS protocol is that policy enforcement functionalities are not supported in the case of streaming data, and as a result in this implementation. %However, a possible future existing architecture could be extended in the future to include this feature.

\section{Conclusion} \label{sec:conclusion}
In this study, we showcased a novel approach for the integration of \ac{IDS} Connector components into the ETSI \ac{MEC} Architecture, in order to address emerging needs for secure, trustworthy and sovereign data exchange. The proposed architecture provides the capability of utilizing and composing Dataspace-enabled MEC Applications, without human involvement. 

To corroborate the usability of this proposed framework, we implemented a cloud-native prototypical solution of our architecture and performed an experiment, comprising of a data exchange use case, with varying data sizes, and compared the time needed for a complete data exchange cycle, both in the case of \ac{IDS} enabled \ac{MEC} applications and in the direct \ac{MEC} App communication scenario. Our findings highlight the feasibility and usability of incorporating an \ac{IDS} Connector-as-a-Service within a \ac{MEC} Platform scheme and provide evidence for the scalability potential of such frameworks.


\section*{Acknowledgment}
This work is part of the European Union’s Horizon Europe research and innovation programme under grant agreement No 101057527 (NextGEM) funded by the European Union. Views and opinions expressed are however those of the authors only and do not necessarily reflect those of the European Union. Neither the European Union nor the granting authority can be held responsible for them.



%\bibliographystyle{IEEEtran}

\end{document}